# When Optical Microscopy Meets All-Optical Analog Computing: A Brief Review

Yichang Shou, Jiawei Liu, Hailu Luo[†]

Laboratory for Spin Photonics, School of Physics and Electronics, Hunan University, Changsha 410082, China

Corresponding author. E-mail: [†]hailuluo@hnu.edu.cn

As a revolutionary observation tool in life science, biomedical, and material science, optical microscopy allows imaging of samples with high spatial resolution and a wide field of view. However, conventional microscopy methods are limited to single imaging and cannot accomplish real-time image processing. The edge detection, image enhancement and phase visualization schemes have attracted great interest with the rapid development of optical analog computing. The two main physical mechanisms that enable optical analog computing originate from two geometric phases: the spin-redirection Rytov-Vlasimirskii-Berry (RVB) phase and the Pancharatnam-Berry (PB) phase. Here, we review the basic principles and recent research progress of the RVB phase and PB phase based optical differentiators. Then we focus on the innovative and emerging applications of optical analog computing in microscopic imaging. Optical analog computing is accelerating the transformation of information processing from classical imaging to quantum techniques. Its intersection with optical microscopy opens opportunities for the development of versatile and compact optical microscopy systems.

## Contents

## 1 Introduction

Optical microscopy is an important tool in observing living cells, imaging biological tissues and life science research [1]. Early improvements in optical microscopy centered on increasing the contrast between the signal and background. In the 1930s, Zernike designed the phase contrast microscopy to convert the relative phase differences of samples into amplitude differences to produce high-contrast imaging [2]. Then, Georges Nomarski improved the Wollaston prism in 1955 to form the differential interference contrast (DIC) microscopy with relief effect [3]. On the other hand, fluorescence microscopy allows imaging of single molecule using the excitation and emission of fluorophores compared to microscopies based on light wave properties [4]. However, due to the optical diffraction limit, the optimization of modern optical microscopy focuses on providing higher spatial resolution [5].

At the beginning of the 21st century, fluorescence super-resolution methods saw an explosive development [6]. Stimulated emission depletion microscopy by Hell [7] in 2000 was the first far-field super-resolution technique for observing live cells. It uses spatial modulation and saturation jumps between two molecular states to break the diffraction barrier. In 1995, Betzig [8, 9] proposed the theoretical idea of super-resolution imaging based on single-molecule precise localization, and experimentally realized photoactivated localization microscopy in 2006. Subsequently, Zhuang [10] proposed stochastic optical reconstruction microscopy in the same year. The imaging principle is that the precise positions of only a few fluorescent molecules are recorded in each scan, and then the entire image is reconstructed through by imaging cycles. In addition, structured illumination microscopy [11] and multiphoton microscopy [12] also provide powerful support to observe the three-dimensional (3D) microstructure of samples.

Optical analog computing provides a new application platform for information processing with its natural properties of low loss and high speed in real time [13-15]. In recent years, optical analog computing has developed various functions such as integration [16,17], differentiation [18-20], solving differential equations [21] and integral equations [22]. Among them, differential operations have important applications in all-optical image processing [23], biomedical diagnostics [24-27], and quantum imaging [28, 29]. All-optical differentiators based on metasurfaces [30-35] and conventional optical interfaces [36-39] have been widely reported. Here, we focus on the two most prominent physical mechanisms for implementing differential operations: the spin-redirection Rytov-Vlasimirskii-Berry (RVB) phase and the Pancharatnam-Berry (PB) phase. Both are geometric phases related to the spin-orbit interaction (SOI). The RVB phase is attributed to the evolution of the beam transport, which produces the conventional photonic spin Hall effect (SHE) at the optical interface [40]. The PB phase is related to the manipulation polarization states of light, and can be used to design the PB phase metasurfaces to generate strong photonic SHE [41, 42].

Despite the rise of many novel microscopy technologies, there is still need to overcome the shortcomings of imaging speed and photodamage in live biological samples [43]. With the development of optical analog computing, the combination with optical microscopy will further extend image processing capabilities and compensate for observation defects [Fig. 1]. The introduction of optical analog computing in optical microscopy can enhance the optical imaging process and provide label-free, high-resolution, edge extraction advantages for samples [44-48]. The multi-mode switching of bright-field, edge, and phase contrast are adjusted by optical analog computing, also laying the foundation for the compact high contrast optical microscopy [49].

In this review, we aim to provide a simple framework for the development and innovation of introducing optical analog computing in optical microscopy. Firstly, the basic theories of the RVB phase and PB phase are explained. These two Berry phases induced photonic SHE is a major branch of performing optical analog computing. Next, we review the application of optical differential operations based on the RVB phase and PB phase in all-optical image processing. We then highlight advances in combining optical analog computing with optical microscopy, including various image processing capabilities in both classical and quantum microscopy. Finally, we summarize the emerging technologies arising from the intersection of these two fields and discuss the future opportunities.

## 2 The spin-redirection RVB phase and PB phase

The geometric Berry phase is a classical representation of the SOI of light between spin angular momentum and orbital angular momentum [50, 51]. In 1984, Berry showed that the cyclic and adiabatic evolution of quantum states in parameter space leads to geometric phases [52]. Inspired by this, the optical similar study of Berry phase was induced, i.e., the spin-redirected RVB phase and PB phase [53, 54]. These two geometric phases correspond to the two different photonic SHE, both of which can perform optical differentiation operations [55]. The former causes a very weak spin splitting, which is generally considered as photonic SHE, by the RVB phase gradient in the momentum space leading to the real space displacement. The latter causes a large SHE displacement, resulting from the PB phase gradient in real space leading to the displacement in momentum space.

When a vector moves parallel along a closed-loop path on a 3D surface back to the starting point, it is rotated by a small angle in the direction compared to the origin. This behavior is called the overall change of the vector. Meanwhile, the vector is not rotated with respect to the normal of the surface during the parallel movement to each adjacent point, i.e., the vector does not change locally [56]. This phenomenon of local invariance and overall change is a geometric phenomenon. In the optical systems, the 3D curved space can also be the momentum space and Stokes parameter space, which are used to describe the RVB phase and PB phase, respectively. When the light propagates at an optical interface, its propagation trajectory changes. It is the wave vector moving geometrically parallel on the sphere of momentum space [Fig. 2(a)]. The polarization vector does not rotate locally, but produces an overall change in the entire surface space, which necessarily induces the RVB phase associated with circularly polarized light. Another Stokes parameter space can be used to describe the spatial evolution of the polarization states [Fig. 2(d)]. The inevitable rotation between the transport vector and global spherical coordinates results in PB phases. In summary, the geometric phases in optics are attributed to the coupling between the angular momentum and rotation of the parametric spatial coordinates [57].

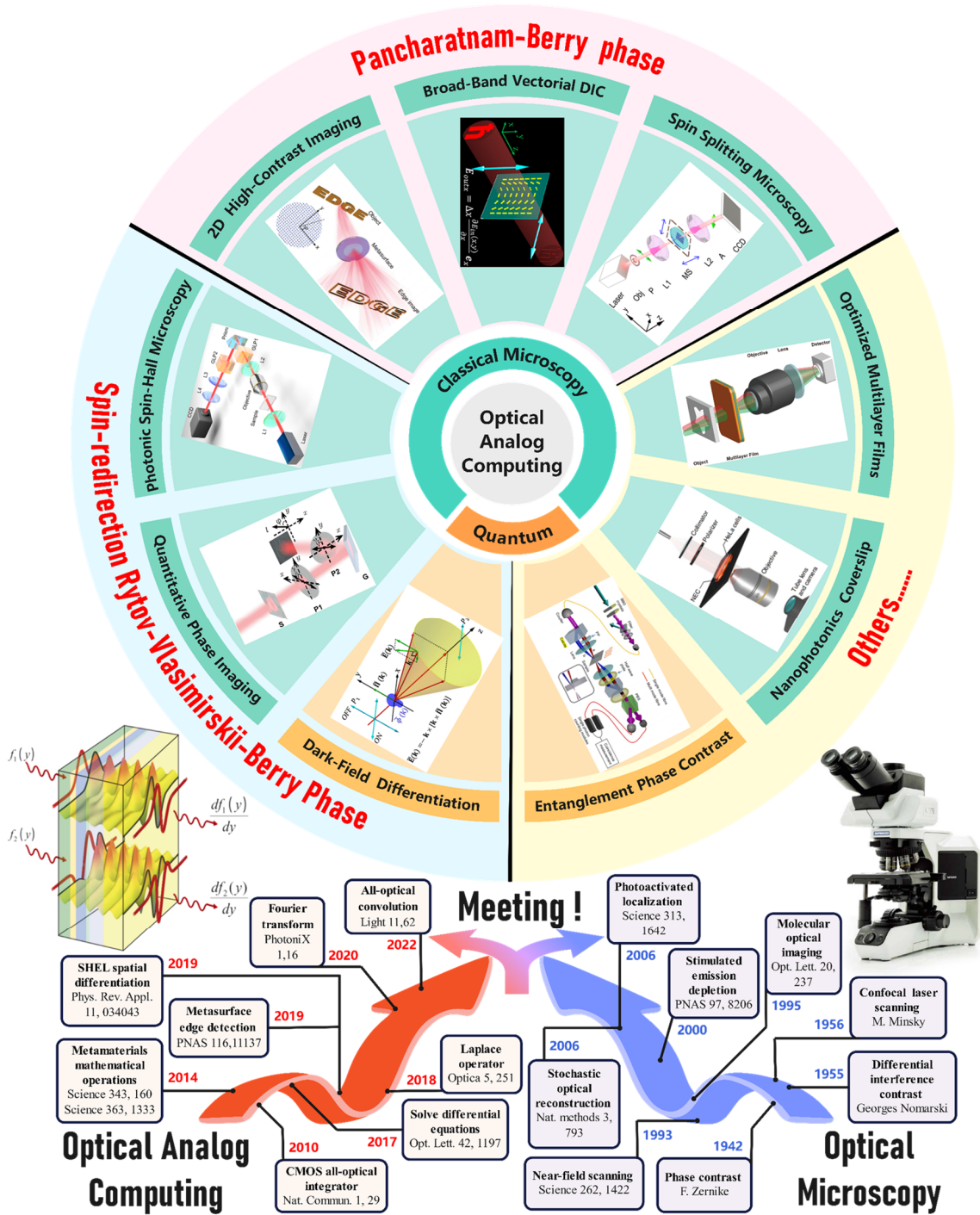

**Fig. 1** An overview of the combination with optical analog computing in optical microscopy, containing the brief development routes of both. [14], Copyright 2014, The American Association for the Advancement of Science. Spin-redirection Rytov-Vlasimirskii-Berry phase: dark-field differentiation [29], Copyright 2022, American Physical Society, quantitative phase imaging [44], under a Creative Commons Attribution (CC BY 4.0) license, photonic spin-Hall microscopy [48], Copyright 2022, American Physical Society. Pancharatnam-Berry phase: 2D high-contrast imaging [26], under a Creative Commons Attribution License, broda-band vectorial DIC [47], American Chemical Society, spin splitting microscopy [83], Copyright 2022, American Physical Society. Others: optimized multilayer films [46], Copyright 2022, John Wiley, and Sons, nanophotonics coverslip [49], Copyright 2021, under the Creative Commons CC BY license, entanglement phase contrast [91], Copyright 2013, Springer Nature.

## 2.1 Spin-redirection RVB phase

The spin-redirection RVB phase was first studied by Rytov and Vladimirskii [58] and then redeveloped in the 1990s. When a paraxial beam is reflected or refracted at an optical interface, the polarization vectors of many plane waves (i.e., angular spectral components) are rotated differently to meet the transverse properties of the electromagnetic field [56]. The resulting RVB phase gradient appears as the spin-dependent shift $\Delta r$ in real space:

$$\Delta r = r_x + r_y = \nabla \Phi_{RVB}(k_x, k_y). \quad (1)$$

$\Phi_{RVB}(k_x, k_y)$ is the $k$-dependent RVB phase. When light is reflected on the optical interface, the $z$-direction is assumed to be the incident direction, where $xz$-plane is the incident plane [59]. The geometric phase gradient along the $k_y$-direction leads to a transverse shift of beam along the $y$-direction [Fig. 2(b)]. In this case, the $k_y$-dependent RVB phase induces spin-dependent splitting vertical to the plane of incidence, i.e., photonic SHE [60]. The RVB phase can be expressed as $\Phi_{RVB}(k_y) = \sigma k_y \delta$ with $\sigma$ refers to the spin state of incident light, and $k_y$ is the transverse wave vector component. $\delta$ is related to the optical parameters of this interface material, for the air-glass interface it is $\delta \propto (1 + r_s/r_p) \cot \theta_i / k_0$ [61]. Where $\theta_i$ is the incident angle, $k_0$ is the wave number in a vacuum, $r_p$ and $r_s$ are the Fresnel reflection coefficients. The polarization states change experienced by photonic SHE can be analyzed by Fig. 2(c). The wave vector moves parallel in momentum space and is always tangent to the $k$-sphere, resulting in different polarization rotations for different angular spectral components of the beam [48].

## 2.2 PB phase

The PB phase was first discovered in the optical field by Pancharatnam [50] in 1956 and popularized by Berry [62] in 1987. When the beam is incident in a birefringent crystal, the direction of crystal optical axis determines the additional phase introduced, i.e., the PB phase. The Jones matrix transformation can be expressed as [42]

$$\begin{bmatrix} 1 \\ \sigma_{\pm i} \end{bmatrix} \to \cos\frac{\psi}{2}\begin{bmatrix} 1 \\ \sigma_{\pm i} \end{bmatrix} + i\sin\frac{\psi}{2}\begin{bmatrix} 1 \\ \sigma_{\mp i} \end{bmatrix} e^{i2\sigma_{\pm}\alpha(x,y)}. \quad (2)$$

The PB phase can be written as $\Phi_{PB}(x, y) = -2\sigma_{\pm}\alpha(x, y)$, where $\alpha(x, y)$ is the local optical axis direction of the medium. $\psi$ is the phase retardation of the crystal. According to the above equation, the incident photons in part $\sin^2(\psi/2)$ reverse their spin direction to acquire anti-phase gradient, while other photons remain unchanged [Fig. 2(e)]. In this process, the coordinate-dependent PB phase produces a geometric phase gradient along the medium surface, where this phase gradient is described as a SHE displacement in momentum space [42]:

$$\Delta k(k_x, k_y) = \nabla \Phi_{PB}(x,y) = \frac{\partial \Phi_{PB}}{\partial x}\hat{e}_x + \frac{\partial \Phi_{PB}}{\partial y}\hat{e}_y. \quad (3)$$

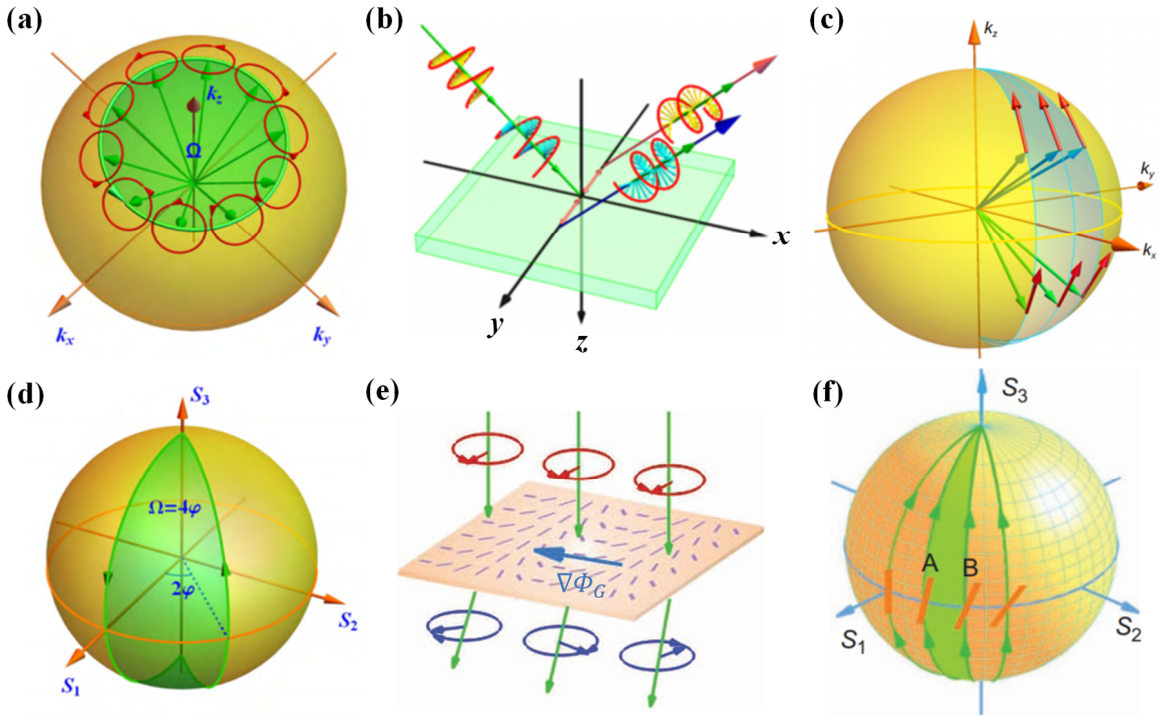

**Fig. 2** Generation of geometric phases and photonic SHE. **(a)** Non-trivial parallel transport of wave vectors in momentum space generates geometric phase. **(b)** Photonic SHE based on the RVB phase at an air-glass interface. **(c)** Geometric features of polarization evolutions on the momentum space. **(d)** Stokes vector for non-trivial parallel transfer in Stokes parameter space. **(e)** Circularly polarized light acquires PB phase gradients through a localized optical axis variation of the metasurface. **(f)** The PB phase-induced polarization evolution on the Poincaré sphere. **(b, c)** Reproduced from Ref. [48], Copyright 2022, American Physical Society. **(e, f)**, Reproduced from Ref. [42], under a Creative Commons Attribution-NonCommercial-NoDerivs 3.0 Unported License.

The Poincaré sphere in Stokes parameter space can better describe the evolution of light polarization states caused by PB phase [Fig. 2(f)]. Light with negative spin evolves to positive spin along different paths from the South Pole to the North Pole. The half of stereo angle surrounded by the green area indicates the corresponding PB phase.

## 3 Optical Analog Computing

As the most developed function in optical analog computing, differential operations play a crucial role in image detection and pattern recognition [63]. Mathematically, differential operation is a linear description of the local rate of change for a function. The object edge is usually where the amplitude, phase, and polarization gradients are large, and the differential operations on the light field can extract the edge information.

### 3.1 Differential operation induced by the RVB phase

The RVB phase induced photonic SHE can be used to perform differential operations when the beam is reflected at a simple optical interface in real space. Photonic SHE leads to spin separation of the left and right circularly polarized (LCP and RCP) components in opposite directions, i.e., transverse displacement $\Delta y$. The reflected electric field can be expressed as:

$$E_r(x,y) = E_{in}(x, y + \Delta y)\begin{bmatrix}1\\-i\end{bmatrix} + E_{in}(x, y - \Delta y)\begin{bmatrix}1\\i\end{bmatrix}. \quad (4)$$

Then, the reflected light passes through a polarizer aligned along the y-axis, and the differential relationship between the output and input field in the spatial domain can be obtained:

$$E_{out}(x,y) = E_{in}(x, y + \Delta y) - E_{in}(x, y - \Delta y)$$
$$\approx 2\Delta y \frac{\partial E_{in}(x,y)}{\partial y}. \quad (5)$$

The earliest experimental work using photonic SHE for one-dimensional (1D) differentiation was done by Zhu et al. [37] who demonstrated that the spatial differentiation is a natural effect of photonic SHE on reflections from any optical interface. The general principle of spatial differentiation is shown in Fig. 3(a), where the linearly polarized light incident on the air-glass interface produces photonic SHE. The reflected LCP and RCP components undergo reverse transverse shift. And output optical field is polarized by a polarizer to eliminate the common linearly polarized part, leaving the circularly polarized part to extract the edge of amplitude object [Fig. 3(b)].

Subsequently, He et al. [38] focused on whether the optical spatial differential at simple optical interfaces is wavelength-dependent. They proposed an optical full differentiator based on the SOI of light. The differentiator is implemented on a glass plate and two polarizers are orthogonal to each other to satisfy the full differentiation of the input optical field [Fig. 3(c)]. As a result, the images edge detection of the $532\ nm$ laser is shown in Fig. 3(d). Due to the pure geometric nature of the SOI of light, the differential effect is light-free with respect to wavelength. At the same time, 1D edge images in any direction can be obtained by adjusting the polarization state of incident beam.

In practical image processing, the objects usually contain amplitude and phase information. Wang et al. [48] found that photonic SHE can perform spatial differentiation of phase distribution and obtain high-contrast images of transparent phase objects. The pure phase objects visualization scheme they developed requires only a simple air-glass interface for reflection [Fig. 3(e)]. For a phase object with phase distribution $\varphi(x,y)$, the output electric field undergoes only phase transformation $E_{out}(x,y) = exp[i\varphi(x,y)]\ e_x$. Then, the light field $I_{out}(x,y) = |exp[i\varphi(x,y)]|^2 = 1$ does not acquire contrast and image is submerged in the laser background. However, after the differentiation system of photonic SHE, the output electric field can be obtained as $E_{out}(x,y) \propto \Delta x\ \partial\varphi(x,y)/\partial x\ exp[i\varphi(x,y)]\ e_x$ and the corresponding optical field is converted to $I_{out}(x,y) \propto |\Delta x\ \partial\varphi(x,y)/\partial x|^2$. As a result, the phase gradient can be transformed into intensity contrast by spatial differentiation operations [Fig. 3(f)]. This interesting scheme has great potential for microscopic imaging of transparent cells and may also be applicable to single-photon imaging.

The photonic SHE not only simply performs differential operations, but can also be used to calibrate the global phase response and evaluate phase delay from two intensity images [64]. Further, the SOI of light-based physical effects for implementing differential operations at optical interfaces include geometric spin Hall effect of light [65], Goos–Hänchen effect [39], and Brewster effect [66, 67]. Unlike conventional photonic SHE, the spin splitting in geometric spin Hall effect is purely a geometric effect, independent of the nature of the interfacial material. The two-dimensional (2D) differential operation brought by the Brewster effect can achieve efficient full edge detection. Similarly, the construction of topological charges in the transfer function to achieve a broadband isotropic 2D differentiator is demonstrated [68].

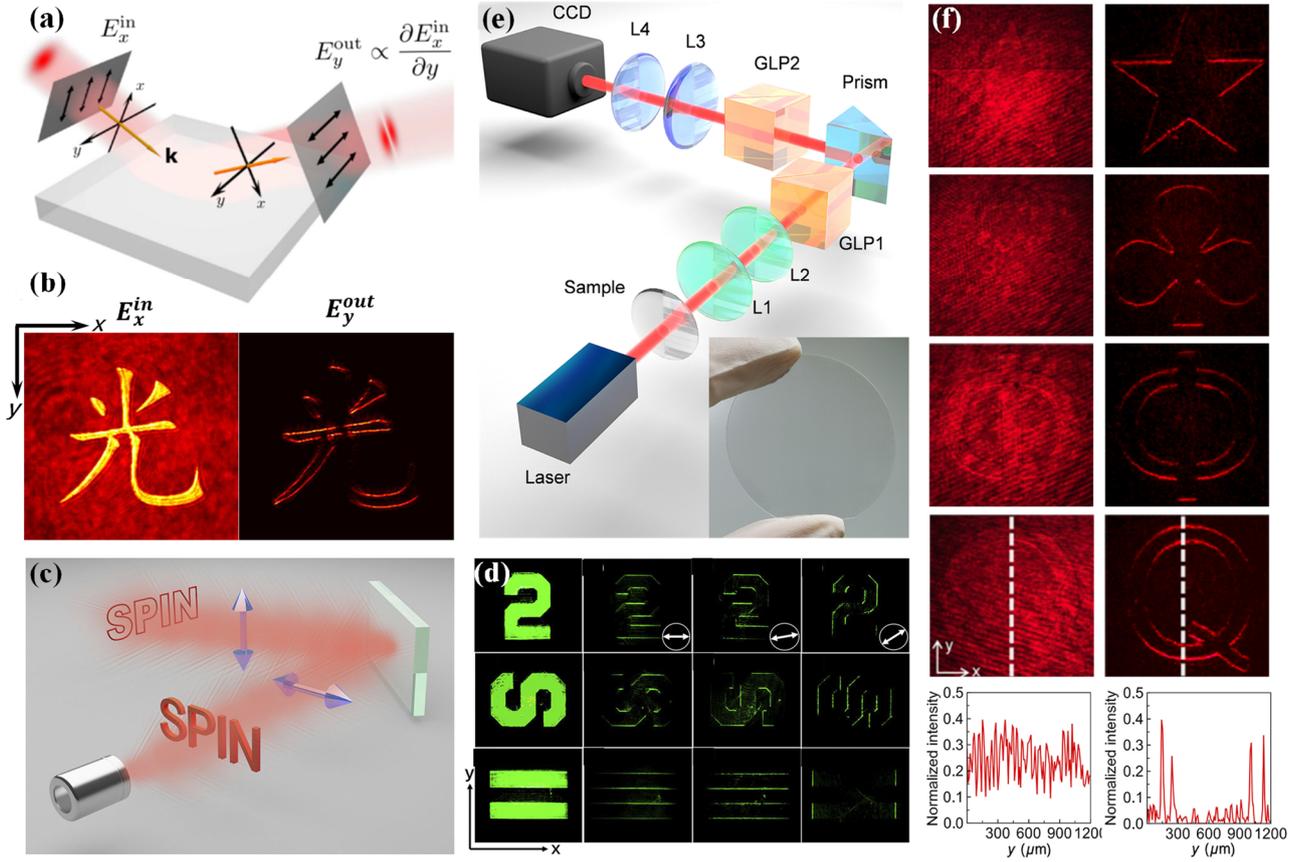

**Fig. 3** The RVB phase-based optical differential operations and image edge detection. **(a)** Schematic illustration of spatial differentiation of photonic SHE at the interface of isotropic media. **(b)** Experimental results of edge detection for amplitude patterns. **(c)** Air-glass reflection interface realization optical fully differentiator. **(d)** The edge detection images when the polarization angle of the incident light is 0°, 10°, 40° respectively. **(e)** Diagram of the photonic spin-Hall differentiator experimental setup, inset shows the actual phase pattern. **(f)** Bright-field and differential images of phase patterns. **(a, b)** Reproduced from Ref. [37], under a Creative Commons Attribution 4.0 International License. **(c, d)** Reproduced from Ref. [38], under a Creative Commons Attribution 4.0 International License. **(e, f)** Reproduced from Ref. [48], Copyright 2022, American Physical Society.

## 3.2 Differential operation induced by the PB phase

The development of metamaterials and metasurfaces provide a path for optical components miniaturization and multi-dimensional manipulation of optical fields [69]. The specially designed metasurface optical axis structure can create a coordinate-dependent PB phase, which induces the constant momentum shift. Consider a linearly polarized light along the $x$-direction incident on the PB phase metasurface (optical axis rotation rate in the $y$-direction), the transmitted field is

$$\tilde{E}_t(k_x, k_y) = \tilde{E}_{in}(k_x, k_y - \Delta k_y) \begin{bmatrix} 1 \\ -i \end{bmatrix}$$
$$+ \tilde{E}_{in}(k_x, k_y + \Delta k_y) \begin{bmatrix} 1 \\ i \end{bmatrix}. \quad (6)$$

Momentum shift $\Delta k_y$ is converted to shift $\Delta y = \sigma z \Delta k_y / k$ in real space after transmission distance $z$. The transmitted light with spin splitting passes through the polarizer (polarization axis along the $y$-direction) to complete the differential operation.

This PB phase induced differential operation described above was implemented in the study of Zhou et al. [20] who designed a high-efficiency dielectric metasurface to perform 1D edge detection. After the linearly polarized light is incident on the metasurface, the LCP and RCP light acquire additional phases of $+2\varphi$ and $-2\varphi$ ($\varphi(x,y) = \pi x/\Lambda$, $\Lambda$ is period) and has a slight transverse shift in image plane to provide edge information [Fig. 4(a)]. The introduced phase delay originates from the geometric phase of the etched nanostructure and is essentially wavelength independent. The experimental results of broadband 1D edge detection are shown in Fig. 4(b). Note that the quality and efficiency of optical differentiation operation results are affected by the metasurface structure. Recently, Xu et al. proposed a method that can reverse the design of PB phase metasurfaces [70]. By deriving the evolution of PB phase on the Poincaré sphere, the relationship between the local optical axis variation and phase can be obtained [Fig. 4(c)]. The all-optical image edge detection optical system used to verify the inverse design metasurface is shown in Fig. 4(d).

High-efficiency metasurfaces make it possible to use multiple metasurfaces to perform 2D and higher-order differential operations in optical system. Wang et al. proposed a broadband vector differentiator based on computing metasurfaces to obtain vector 2D and high-contrast edge information [47]. The optical system they designed contains two identical metasurfaces placed in two paths of the Mach-Zehnder interferometer [Fig. 4(e)]. The metasurface on path 1 performs 1D differential operations in the $x$-direction, and another on path 2 performs differential operations in $y$-direction by rotating the optical axis direction. The 1D edge images can be observed when the incident beam illuminating the intensity object passes through only one path, while 2D edge detection is achieved when the differential operations in the $x$- and $y$-directions converge [Fig. 4(f)].

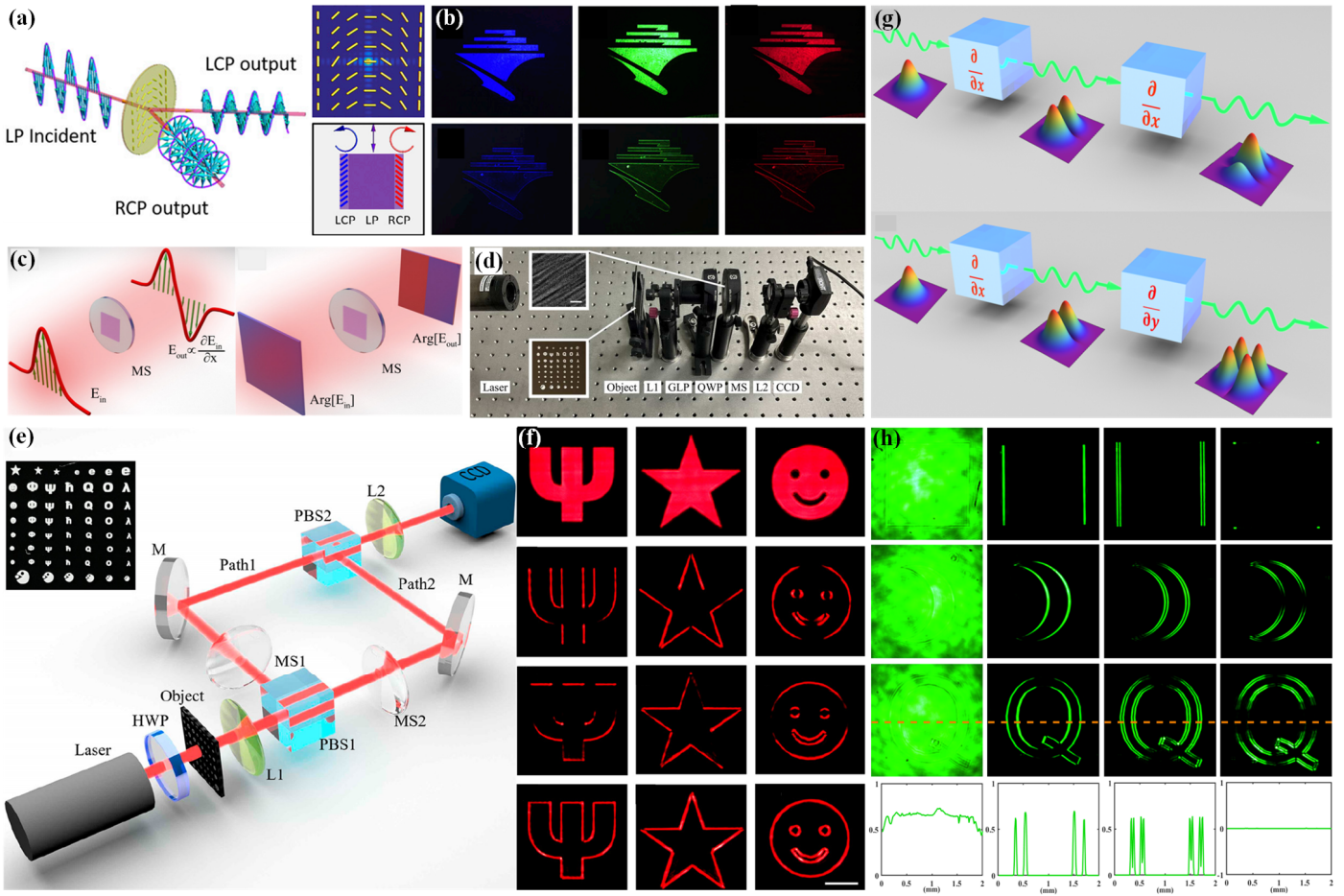

**Fig. 4** The PB phase-based optical differential operations and image edge detection. **(a)** PB phase metasurface differentiator. The right figures represent the metasurface-induced Fourier space spectrum and real-space image. **(b)** Broadband edge detection experiments with three wavelengths of $430\,nm$, $500\,nm$, and $670\,nm$, respectively. **(c)** Optical differential operations and corresponding phase distributions based on inversely designed metasurface. **(d)** All-optical image edge detection optical system. **(e)** Experimental setup for vector differential operations based on computational metasurface. The inset presents the resolution target (object). **(f)** 1D differential operations in $x$- and $y$-directions and 2D edge detection images. **(g)** Schematic diagram of a higher-order spatial differentiator based on cascaded operations. **(h)** All-optical spatial second-order differentiator for phase objects. **(a, b)** Reproduced from Ref. [20], under a Creative Commons Attribution-Noncommercial NoDerivatives License 4.0. **(c, d)** Reproduced from Ref. [70], Copyright 2022, under an exclusive license by AIP Publishing. **(e, f)** Reproduced from Ref. [47], Copyright 2022, American Chemical Society. **(g, h)** Reproduced from Ref. [71].

Unlike the parallel 2D differentiation operation using the Mach-Zehnder interferometer, cascaded higher-order spatial differentiators have also been proposed recently [71]. To simulate the cascaded operation of electronic system, two first-order differentiators that can perform different directions are cascaded to achieve second-order differentiation [Fig. 4(g)]. The image results of the second-order differentiation processing phase object are shown in Fig. 4(h). First-order differential edge images are then subjected to second-order differentiation, and images present double edge effect. Other analog units can be inserted in the cascaded system to realize optical analog computing networks in any combination according to the actual needs.

In addition to the use of metasurfaces for optical analog computing, liquid crystal flat lens based on PB phase fabrication can also achieve 2D edge detection with up to 97% transmission efficiency [72]. Metasurfaces designed using other principles such as nonlocal, quasi-bound states, and Mie-resonant are also suitable for various image processing scenarios [73-75]. These optical components may have important applications in high contrast microscopes and compact optical systems.

## 4 Optical analog computing towards microscopy imaging

Imaging without staining labels and with reduced phototoxicity is one of the current pursuits of optical microscopy [43, 76]. Optical analog computing based on the RVB and PB phase have the advantages of non-fluorescent staining, fast, and high resolution. By introducing highly compatible optical analog operations into microscopic imaging, the geometric features of biological samples can be extracted for edge enhanced. This will further extend the utility of optical microscopy for in vivo imaging and contamination-free histopathology.

### 4.1 Classical microscopy

#### 4.1.1 Introducing RVB phase to optical microscopy

The RVB phase-based spatial differentiation at optical interfaces and phase enhanced imaging do not depend on complex structures. Combining the RVB phase with optical microscopy offers the advantage of low cost and compactness. Edge detection and recovery of phase distribution can be achieved with only one glass interface, avoiding the difficulties of optical alignment, as the case of Brewster differential microscopy [77]. This simultaneously provides great simplicity and flexibility for developing multifunctional microscopes.

In conventional microscopy, phase contrast makes the transparent cells visible and DIC provides phase gradient of the sample, but neither can quantify the phase distribution. Zhu et al. [44] demonstrated a tunable spatial differentiator to implement phase mining based on the single dielectric interface light reflection. The invisible phase structure of the cells is transformed into a structured contrast image after polarization modulation analysis of reflected light through the air-glass interface [Fig. 5(a)]. By introducing a small deviation value $b$ as constant background, the differential contrast of phase objects can be enhanced. The first-order derivatives of the phase distribution in the *x*- and *y*-directions are shown in Fig. 5(b), with the epithelial cells in relief. The phase results are recovered from Fig. 5(b) using the 2D Fourier algorithm, which is highly consistent with the original phase distribution [Fig. 5(c)].

The spin splitting phenomenon of RVB phase induced photonic SHE reflected at optical interface is very interesting. Wang et al. [48] proposed a compact differential microscope based on photonic SHE for differential imaging and quantitative phase analysis of pure phase samples. The optical path for differential microscopic imaging at a simple glass interface is shown in Fig. 5(d). The incident beam strikes the prism-air interface at an angle of 45° and the reflected beam passes through GLP2 to analyze the output polarization state. Unstained onion and shallot cells are almost difficult to distinguish cell structures in bright-field images due to their pure phase characteristics. On the other hand, when passed through optical differential microscopy, the cell outlines were visualized as high-contrast edge images [Fig. 5(e)].

It is worth noting that introducing appropriate bias delays in differential scheme can reconstruct the phase distribution of pure phase objects. Experimentally, GLP1 and GLP2 are orthogonal to output edge images. The introduction of the bias delay is the basis for rotating the polarization axis $\pm 0.8°$ of GLP2 to obtain the distributions of $\partial\varphi(x,y)/\partial x$ or $\partial\varphi(x,y)/\partial y$. The reconstructed phase distribution of the focal star with $350\ nm$ phase gradient is shown in Fig. 5(f). The images of partial derivatives in the *x*- and *y*-directions are vector superimposed, and results obtained are then used to recover the phase distribution by 2D Fourier integration. Quantitative phase imaging (QPI) quality is determined by reconstructing the intensity profile of the phase distribution (red line), consistent with the phase gradient distribution (blue line).

It should be noted that QPI is an important imaging technique in the biomedical field that has attracted much attention recently. In addition to the above-mentioned quantitative phase analysis accomplished using an optical interface, the single-shot QPI achieved by two cascaded metasurfaces has been proposed [78]. The metasurfaces based approach promotes a fast and compact configuration for integration into commercial optical microscopes [79]. Common to these solutions is the capture of two or three images for computation to produce quantitative phase gradient images.

#### 4.1.2 Introducing PB phase to optical microscopy

Metasurfaces can manipulate the electromagnetic properties of light with a large degree of freedom [80]. Traditional control of optical arbitrary wave fronts relies on the physical effects accumulated by various optical elements during the propagation of light, while complex optical functions can be achieved by a single metasurface. Specially designed metasurfaces have been shown to be compatible with optical devices and to improve imaging contrast [81, 82].

In 2021, Zhou et al. [26] further proposed a broadband 2D spatial differentiator based on a PB phase dielectric metasurface. The tailored metasurface has a symmetric phase gradient in the radial direction, i.e., the LCP and RCP are radially separated to ensure 2D differentiation. The differential scheme they designed enables edge detection of intensity and phase samples by simply inserting this metasurface into the commercial microscope optical path [Fig. 6(a)]. The light scattered by cells is collected and imaged by the objective lens, with the metasurface placed on the back focal plane of the objective lens. The different microscopic observations of human umbilical vein endothelial cell (HUVEC) are shown in Fig. 6(b). Bright-field images have almost no visible cell features, and dark-field with phase contrast simply see the structure of cells. And after inserting the metasurface, the edge detection can extract outline and cell characteristics are clearly visible. This scheme has high sensitivity and accuracy in detecting transparent biological specimens.

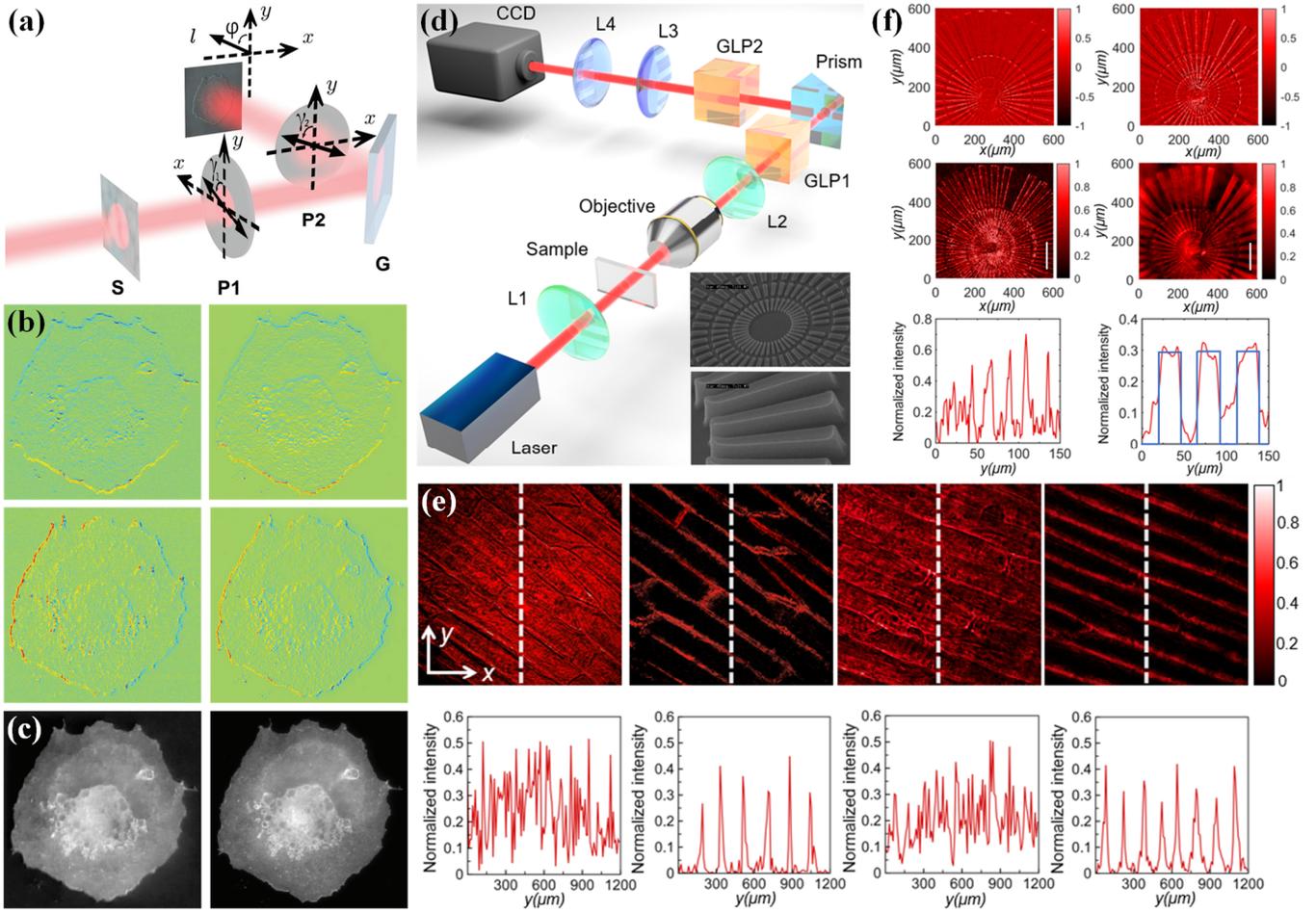

**Fig. 5** Microscopically enhanced imaging based on optical interface reflection. **(a)** Schematic diagram of phase analysis for polarization modulation at the air-glass interface. **(b)** Results of horizontal partial derivatives of epithelial cell phase distribution. **(c)** 2D Fourier algorithm to recover phase results and original phase distribution. **(d)** Photonic spin-Hall differential microscopy optical path. **(e)** Bright-field and differential images of unlabeled onion and shallot cells. **(f)** Quantitative phase microscopy reconstruction of the 350 $nm$ focus star phase distribution. **(a-c)** Reproduced from Ref. [44], under a Creative Commons Attribution (CC BY 4.0) license. **(d-f)** Reproduced from Ref. [47], Copyright 2022, American Physical Society.

In the following year, Zhou et al. [83] proposed another Fourier optical spin splitting microscopy (FOSSM) using the PB phase metasurface. The metasurface is placed in the Fourier plane of the microscopic optical path [Fig. 6(c)], which separates object image into two replicas of the LCP and RCP components. The bias retardance of these two replicas is introduced by the relative position of metasurface or the incident beam polarization direction. The quantitative phase information of the transparent sample is obtained by relying on circularly polarized light interference. Experimentally, FOSSM implements single-shot quantitative phase gradient imaging (QPGI) by translating the metasurface to acquire a series of retardance images with different bias retardation [Fig. 6(d)]. Four subframes of NIH3T3 cells were captured simultaneously using a polarization CMOS camera to distinguish four retardance images with phase delays of 0° to 270°. Then, the phase gradient information of the samples is calculated using the four-step phase shifting method.

DIC microscopy is performed by two Nomarski prisms with the ability to visualize transparent samples, but is limited to 1D imaging. Recently, Wang et al. [47] designed a 2D vector DIC microscopy based on computing metasurfaces. Benefiting from the compactness and versatility of the metasurfaces, Nomarski prism in the DIC microscopy is replaced by the metasurfaces. And unlike the conventional optical path, two computing metasurfaces are placed on each of the two branches of the Mach-Zehnder interferometer to perform 1D differential operations [Fig. 6(e)]. The imaging effect of the optical vector differentiator plugged into bright-field microscope is shown in Fig. 6(f). The resolution target is barely visible in bright-field observation, and partial edge information can be displayed by differentiation operations in the *x*- and *y*-directions. And when they are combined by PBS2, 2D DIC images are presented. The vector differentiation based on metasurfaces optimizes conventional DIC functionality, and provides broad-band and high-contrast phase imaging.

Moreover, unlike the LCP and RCP splitting produced by the PB phase metasurfaces. The spiral phase technology is based on radial Hilbert transform filtering, which introduces the phase difference of π between the positive and negative spatial frequencies of light field [25, 45, 84]. Previously, related studies used laser sources, while commercial microscopes used halogen lamps or light-emitting diode (LED). Recently phase contrast microscopy based on vortex topology quadrupole using LED non-coherent light source was also experimentally realized [85].

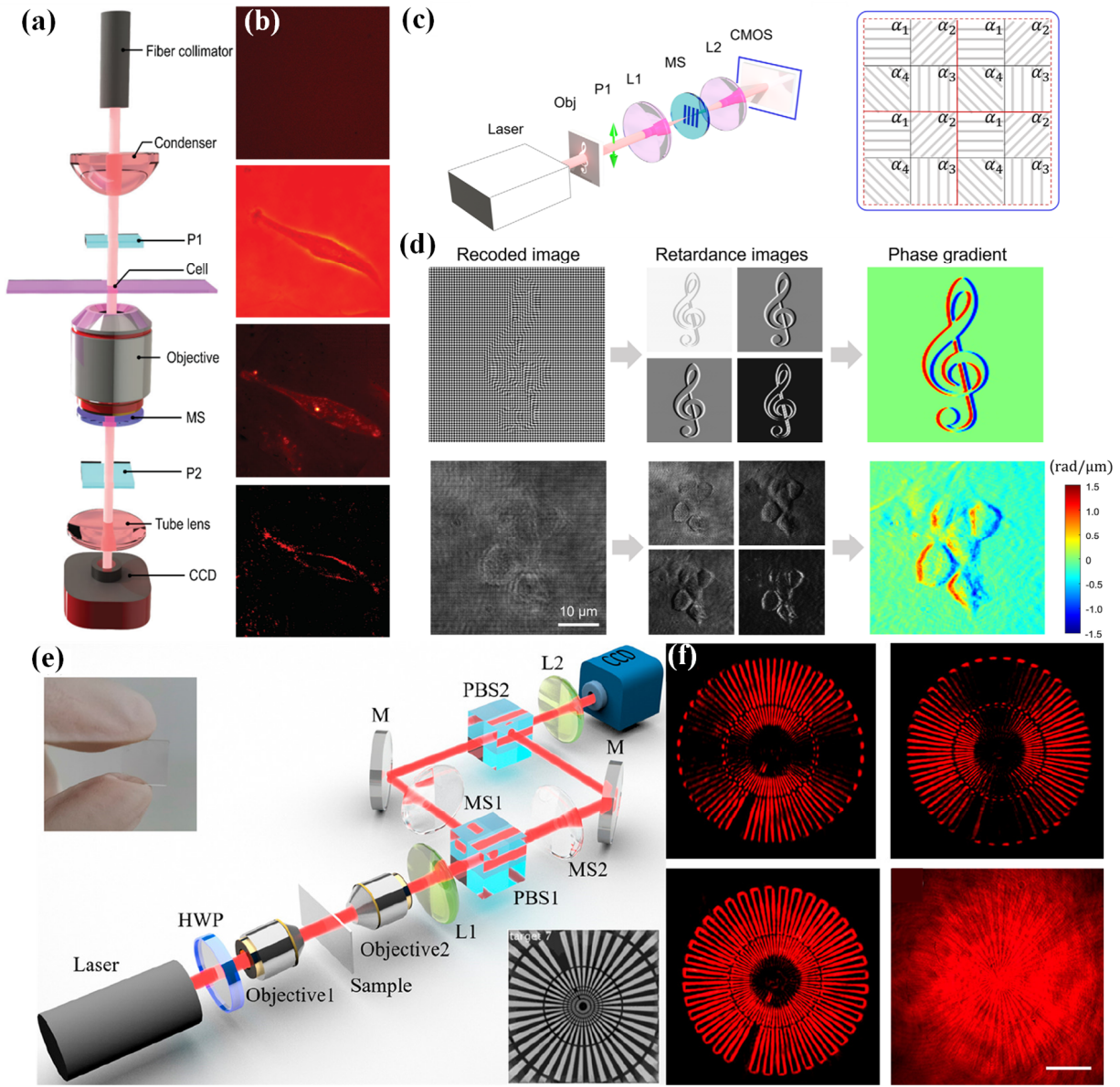

**Fig. 6** High-contrast microscopy imaging based on the PB phase metasurfaces. **(a)** 2D PB phase metasurface combined with commercial microscopy. **(b)** Different microscopic modes of observing HUVEC cells, from top to bottom, bright-field, dark-field, phase contrast, and differential images, respectively. **(c)** Metasurface single shot QPGI optical path and polarization direction of camera pixels. **(d)** Pattern simulation and NIH3T3 cells imaging for single shot QPGI. **(e)** Experimental setup of the vectorial DIC microscope. **(f)** Phase target 1D and 2D edge microscopy images. **(a, b)** Reproduced from Ref. [29], under a Creative Commons Attribution License. **(c, d)** Reproduced from Ref. [83], Copyright 2022, American Physical Society. **(e, f)** Reproduced from Ref. [47], Copyright 2022, American Chemical Society.

## 4.2 Quantum microscopy

The resolution of conventional optical microscopy can be limited by diffraction limits, hindering the exploration of sub-cell structures. Quantum microscopy is generally achieved by using quantum light source illumination in the microscope system, and it is an important tool for characterizing the structure and understanding the dynamics of living systems [86, 87]. Quantum illumination technology was first proposed by Seth Lloyd, where the entanglement photons can be used to improve system sensitivity in the context of a large interference environment [88]. For systems with N-photon entanglement, quantum illumination improves the signal-to-noise ratio (SNR) by a factor of 2N compared to classical systems [89]. Therefore, microscopic imaging using a quantum light source can suppress quantum noise and exceed the standard quantum limit (SQL), providing higher SNR and resolution [90].

One of the first studies that used entangled sources in microscope and systematically discussed SNR was done by Ono et al. [91] in 2013. They designed an entanglement-enhanced microscope that can also be called quantum differential microscope, i.e., using a NOON state light source to illuminate the conventional DIC optical path [Fig. 7(a)]. The NOON state is a two-mode state where N photons are in the superposition state and all photons are in one of the modes, and its entanglement is generally achieved with the help of path or polarization [92]. A comparison of the entanglement-enhanced imaging results is shown in Fig. 7(b). At the same average total number of low photons, the step protrusions of the 2D scan image taken by entangled light are clearly visible, while classical illumination is blurred. SNR exceeded the SQL by a factor of 1.35±0.12 compared to the classical image visibility. The signal intensity of the red box in the image is indicated by the photon compliance number distribution below.

Non-destructive imaging of living biological samples has been a key issue considered in microscopy field. And the indirect reduction of environmental noise by increasing the light intensity largely introduces phototoxicity [76]. The natural low light environment of quantum microscopy provides a suitable observation platform for reducing phototoxicity and achieving high-quality imaging of non-invasive cell tissues. Liu et al. [29] introduced optical analog computing to quantum system and proposed a quantum dark-field microscopy based on intrinsic optical differential. The intrinsic differential operation makes pure phase objects visible at low photon levels, which arise naturally from the cross-polarized components of the dipole scattering field.

An intuitive physical picture of the dipole-emitting electromagnetic field can be described by introducing the Hertzian vector potential [93] [Fig. 7(c)]. The mixed state $1/\sqrt{2}\,(|HH\rangle + |VV\rangle)$ prepared by entanglement source enters the heralding arm and imaging arm without any contact between the two optical paths, and imaging is triggered by conformal measurements only. For most $|H\rangle$ incident photons a bright-field image is formed through the polarizer. While for $|V\rangle$ incident photons a few $|H\rangle$ photons are produced during dipole scattering, and these $|H\rangle$ photons associated with the high $k$-component form dark-field images [Fig. 7(d)].

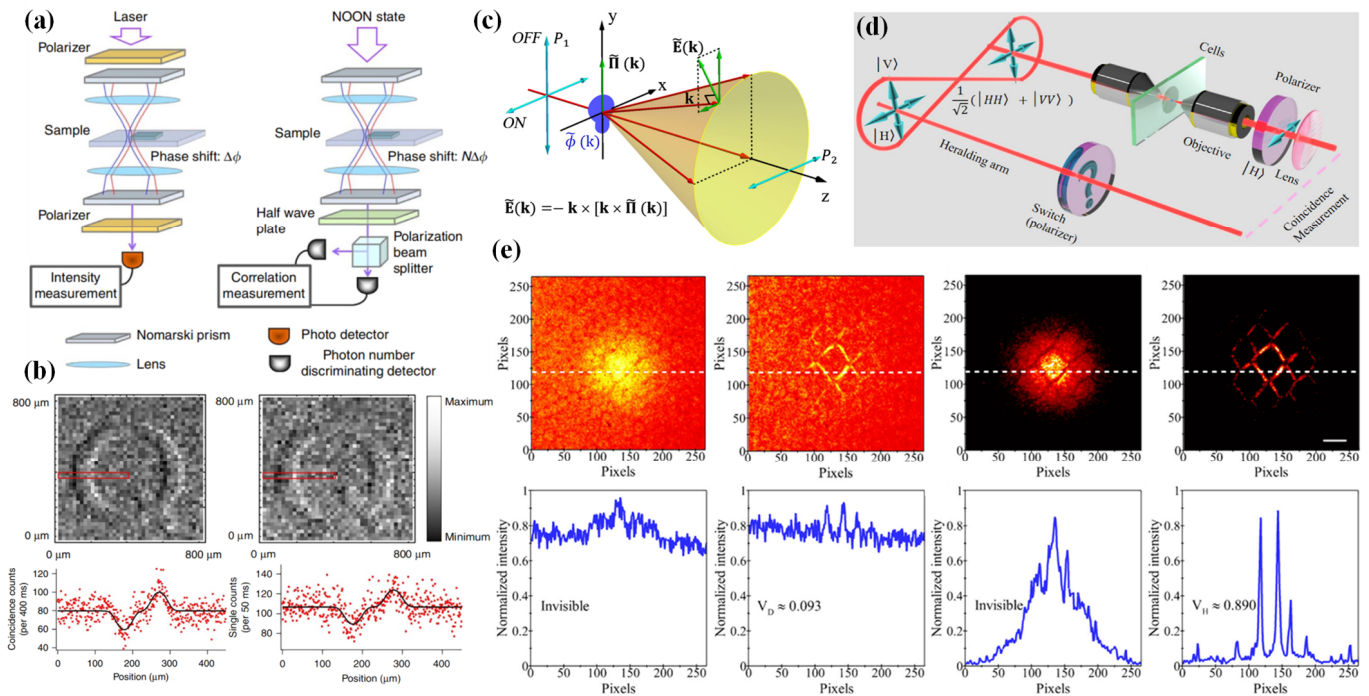

**Fig. 7** Quantum differential and dark-field microscopies. **(a)** How conventional DIC and quantum differential microscopies work. **(b)** Images of sample taken using the entanglement and classical light source illumination. Conformity count data are taken from the red box and the black solid line is the theoretical fit curve. **(c)** The electromagnetic field emitted by an oscillating dipole. **(d)** Schematic diagram of a quantum bright and dark field imaging with remote control of polarization entanglement. **(e)** Bright- and dark-field images of cells taken with ICCD internal and external triggers. **(a, b)** Reproduced from Ref. [91], Copyright 2013, Springer Nature. **(c-e)** Reproduced from Ref. [29], Copyright 2022, American Physical Society.

One of the advantages of quantum imaging lies in herald imaging, which is triggered by an external electrical signal to the ICCD response before receiving the signal photons. Signal photons are detected in a very short optical gating time, thereby rejecting most ambient noise photons, resulting in high contrast quantum images. In contrast, conventional direct imaging signal and ambient noise photons are unconditionally received in integration time. The bright and dark fields of the cells are shown in Fig. 7(e), where direct triggering leads to cell information is drowned out by the environmental noise. While the background noise is minimized in the external trigger, the cell structure is clearly visible and the edge enhancement of the dark-field is remarkable. Their results confirm the finding that herald imaging can be used to obtain high contrast at low photon levels [94]. What needs to be stated is that when high intensity illumination is used, the herald imaging loses its advantage.

The research on the combination of optical analog computing and various imaging technologies extends from the visible spectrum to the ultraviolet and infrared range. In quantum image processing, Padgett et al. [95] used a ghost imaging system to demonstrate that edge images can be used to prove that quantum entanglement violates Bell inequality. Shi [96, 97] group implemented real-time quantum edge-enhanced imaging using spontaneous parametric up-conversion and down-conversion, respectively. In recent years, increasing interest has focused on the field of quantum microscopy imaging [98]. Quantum microscopes such as NOON states, small number of photons, sub-shot-noise, undetected photons, and quantum enhancement techniques have been developed [99-103].

# 5 Summary and outlook

Optical microscopy has penetrated all biology by virtue of its non-contact nature, minimal damage to samples, and abundant imaging mechanisms [104]. Image processing is a key technology in various scientific and engineering disciplines. In the last decade, there has been extensive research on optical analog computing, especially optical differential operations have developed rapidly. In this review, we briefly discuss the rich phenomenon and emerging applications of the cross-fertilization of optical analog computing and optical microscopy. Photonic SHE is one of the tools for implementing optical differentiation operations, which originates from the geometric phases (RVB phase and PB phase) in the SOI of light. Edge detection technology whether it is to identify the intensity object or to observe the phase biological cells in combination with microscope, can realize the edge enhanced high-quality images.

The development of optical microscopy from the 17th century to modern times has given rise to a wide variety of microscopic imaging techniques. Nonlinear near-field optical microscopy allows real-time subwavelength imaging of surface waves [105]. Photonic chip-based super-resolution microscopy analyzes frozen tissue samples [106], and label-free 3D microscopy technique was used to achieve 3D phase-contrast imaging [107]. Quantum imaging is also a hot topic of current interest compared to classical microscopy [108, 109]. Imaging and correct biological interpretation of cancer cells requires unperturbed, and entangled two-photon probes in scanning microscopy allow maximum avoidance of photobleaching at low light intensities [110]. Also in the semiconductor industry, quantum imaging-based infrared microscopy can likewise enhance the imaging quality of silicon chips [111].

Optical analog computing is not only limited to the laboratory, but engineering designs are also interested in wave-based analog computing. By tuning the coupling between the nonlocal response and the scattering harmonics, complex simulations can be realized with small dimensions [112]. The development of all-optical image processing has also led to a series of results, all-optical convolutional and pepper-salt denoising by experimental implementation [27]. Device miniaturization and integration are trends in the development of optical systems. With the advancement of photonic integrated devices, optical operations based on different artificial intelligence models can play a huge role in practical applications [113]. However, real objects often contain 3D information, and it is still a challenge to realize optical analog computing of complex 3D objects [114].

In conclusion, optical microscopy has evolved over the centuries, and the rudiments of optical analog computing can be traced back to Fourier optics, but the two fields have only recently begun to cross-fertilize. Optical microscopy can introduce the broad-band, real-time, and versatile image processing techniques when it meets optical analog computing, extending from classical optics to the quantum field. Despite many advances that has been achieved, the potential of optical analog computing in optical microscopy has not yet been fully exploited. There are still some problems and challenges to be solved, the effective integration of optical analog computing units into commercial microscopes. How the various functions brought by analog operations like differentiation, integration, etc. are integrated to assist in microscopic image processing. Apart from this, the development of quantum analog computing microscopes in the single photon field is also a promising task. Although there is still a long way to go before industrialization and productization, the collision of the two fields will certainly bring new application opportunities.

Acknowledgements This work was supported by the National Natural Science Foundation of China (No. 12174097), the Natural Science Foundation of Hunan Province (No. 2021JJ10008)